\documentclass[11pt]{article} 

\usepackage[usenames,dvipsnames]{xcolor}
\definecolor{Gred}{RGB}{219, 50, 54}
\definecolor{ToCgreen}{RGB}{0, 128, 0}

\usepackage[letterpaper,margin=1in]{geometry}
\usepackage{titling}

\usepackage{wrapfig}
\usepackage{amsmath}

\usepackage{framed}
\usepackage[noend]{algpseudocode}
\usepackage[lined,boxed,ruled,norelsize,algo2e,linesnumbered]{algorithm2e}

\usepackage{amsthm,amsmath,mathtools}
\usepackage{titlesec}
\titleformat*{\paragraph}{\bfseries}

\usepackage[colorlinks]{hyperref}
\hypersetup{
      colorlinks=true,
  citecolor=ToCgreen,
  linkcolor=Sepia,
  filecolor=Gred,
  urlcolor=Gred
  }

\usepackage{empheq}
\usepackage{dsfont}
\usepackage{graphicx}
\usepackage{enumitem}
\usepackage{aliascnt} 
\usepackage{xspace}
\usepackage{booktabs}

\usepackage{algorithm}
\usepackage{multicol}
\usepackage{mdframed}
\usepackage{algpseudocode}

\numberwithin{equation}{section}
\usepackage[tt=false]{libertine}
\usepackage{mathtools}

\usepackage{amssymb,mathrsfs}
\usepackage[varbb]{newpxmath}

\let\savedbigtimes\bigtimes
\let\bigtimes\relax
\usepackage{mathabx} 
\let\bigtimes\savedbigtimes

\usepackage[margin=1in]{geometry}
\usepackage{graphicx}
\usepackage{bbm}
\usepackage{hyperref,color}
\usepackage[capitalize,nameinlink]{cleveref}
\usepackage[dvipsnames]{xcolor}
\hypersetup{
	colorlinks=true,
	pdfpagemode=UseNone,
    citecolor=purple,
    linkcolor=purple,
    urlcolor=black,
	pdfstartview=FitW
}
\usepackage{appendix}
\usepackage{caption}
\RequirePackage{subfigure}

\renewcommand{\Pr}{\mathop{\bf Pr\/}}
\newcommand{\E}{\mathop{\bf E\/}}

\newcommand{\lp}{\left}
\newcommand{\rp}{\right}
\usepackage{pgfplots}

\newcommand{\eps}{\epsilon}

\newcommand{\calY}{\mathcal{Y}}

\def\<{\langle}
\def\>{\rangle}

\def\wh{\widehat}

\def\NP{\mathsf{NP}}
\def\RP{\mathsf{RP}}
\def\APX{\mathsf{APX}}

\crefname{appsec}{Appendix}{Appendices}
\usepackage{tikz}
\usepackage{bm}
\usepackage{mathtools}
\usepackage{color-edits}
\addauthor[Alkis]{ak}{blue}

\newtheorem{theorem}{Theorem}[section]

\newtheorem{lemma}[theorem]{Lemma}
\newtheorem{corollary}[theorem]{Corollary}

\newtheorem{claim}[theorem]{Claim}

\theoremstyle{definition}

\newtheorem*{assumption*}{Assumption}
\newtheorem*{theoremone}{Theorem~\ref{theorem:main}}
\newtheorem*{theoremtwo}{Theorem~\ref{theorem:main2}}
\newtheorem*{lemgadgetwo}{Lemma~\ref{lemma:gadget2}}
\newtheorem{remark}[theorem]{Remark}

\crefname{lemma}{Lemma}{Lemmas}
\crefname{theorem}{Theorem}{Theorems}
\crefname{definition}{Definition}{Definitions}
\crefname{fact}{Fact}{Facts}
\crefname{claim}{Claim}{Claims}
\crefname{proposition}{Proposition}{Propositions}

\newcommand{\SpecIsing}{\mbox{\textsc{SpectralIsing}}}
\newcommand{\BSpecIsing}{\mbox{\textsc{BoundedSpectralIsing}}}

\definecolor{myC}{rgb}{0, 255, 255}
\definecolor{myY}{rgb}{204, 204, 0}
\definecolor{myM}{rgb}{255, 0, 255}
\definecolor{secinhead}{RGB}{249,196,95}
\definecolor{lgray}{gray}{0.8}

\usepackage{appendix}
\crefname{appsec}{Appendix}{Appendices}

\begin{document}
\title{
On Sampling from Ising Models\\
with Spectral Constraints\thanks{
For the purpose of Open Access, the author has applied a CC BY public copyright licence to any Author Accepted Manuscript (AAM) version arising from this submission.}
}
\author{Andreas Galanis \\ University of Oxford\\
\small\textsf{andreas.galanis@cs.ox.ac.uk}
\and
Alkis Kalavasis \\ Yale University \\
\small\textsf{alkis.kalavasis@yale.edu}
\and 
Anthimos Vardis Kandiros \\ MIT\\
\small\textsf{kandiros@csail.mit.edu}
}
\date{}
\maketitle


\begin{abstract}
\small

We consider the problem of sampling from the Ising model when the underlying interaction matrix has eigenvalues lying within an interval 
 of length $\gamma$. Recent work in this setting has shown various algorithmic results that apply roughly when $\gamma< 1$,  notably with nearly-linear running times based on the classical Glauber dynamics. However, the optimality of the range of $\gamma$ was not clear since previous inapproximability results developed for the antiferromagnetic case (where the matrix has entries $\leq 0$) apply only for $\gamma>2$. 

To this end, Kunisky (SODA'24) recently provided evidence that the problem becomes hard already when $\gamma>1$ based on the low-degree hardness for an inference problem on random matrices. Based on this, he conjectured that sampling from the Ising model in the same range of $\gamma$ is NP-hard.

Here we confirm this conjecture, complementing in particular the known algorithmic results by showing NP-hardness  results for approximately counting and sampling when $\gamma>1$, with strong inapproximability guarantees; we also obtain a more refined hardness result for matrices where only a constant number of entries per row are allowed to be non-zero. The main observation in our reductions is that, for $\gamma>1$, Glauber dynamics mixes slowly when the interactions are all positive (ferromagnetic) for the complete and random regular graphs, due to a bimodality in the underlying distribution.  While ferromagnetic interactions typically preclude NP-hardness results, here we work around this by introducing in an appropriate way mild antiferromagnetism, keeping the spectrum roughly within the same range. This allows us to exploit the bimodality of the aforementioned graphs and  show the target NP-hardness by  adapting  suitably previous inapproximability techniques developed for antiferromagnetic systems.
\end{abstract}
\thispagestyle{empty}
\setcounter{page}{0}
\newpage
\section{Introduction}
The Ising model with a symmetric interaction  matrix $J \in \mathbb R^{N \times N}$ is a probability distribution $\mu_J$ over  $\{-1,1\}^N$ with
\begin{equation*}
    \mu_J(\sigma) = \frac{1}{Z_J} \exp\lp(\frac{1}{2}\sigma^\top J \sigma\rp) \mbox{ for all vectors } \sigma\in \{-1,1\}^N,
\end{equation*}
where the normalizing constant $Z_J = \sum_{\sigma \in \{-1,1\}^N} \exp\lp(\frac{1}{2}\sigma^\top J \sigma\rp)$ is the \emph{partition function} of the  model. The most well-studied setting for the Ising model is when the underlying matrix $J$ corresponds to the adjacency matrix of a graph $G$, scaled by a real parameter $\beta$ which corresponds  to the (inverse) temperature;\footnote{Note that in this parametrization $\mu_J(\sigma) \propto \exp(\frac{1}{2}\beta \sigma^\top A \sigma)$, where $A$ is the adjacency matrix of the graph.} for $\beta>0$ the model is called ferromagnetic, and antiferromagnetic otherwise. The more general setting with non-uniform weights in the entries of $J$ arises frequently in statistical learning settings, see, e.g., \cite{learn2,learn3,learn5, learn4}.

The Ising model  is the most fundamental example of a spin system, capturing how local interactions affect the global macroscopic behaviour, see \cite{talagrand2010mean,mezard1987spin,mezard2009information,lauritzen1996graphical} for applications in various areas. From a computer science perspective, sampling from the Ising model plays a key role in various learning and inference problems. Understanding the limits of efficient sampling has therefore been a major focus in the literature, yielding new algorithmic techniques as well as exploring the power of classical algorithms (such as Glauber dynamics) and their connections to phase transitions in statistical mechanics; we briefly review some of the relevant literature below. 

 The prototypical setting where the problem of sampling for the Ising model has been studied is lattices (such as $\mathbb{Z}^2)$, where the landscape for Markov-chain algorithms has been well-understood \cite{martinelli1994approach,martinelli1994approachb,lubetzky2012critical}. Random graph models have also been considered more recently such as sparse random graphs  \cite{ MosselSly,dembo2010ising,fastBD,efthymiou2024sampling,liu2024fast} or the Sherrington-Kirkpatrick  model \cite{eldan2022spectral,el2022sampling,huang2024sampling}.
More closely related to the setting considered in this paper is the case of general graphs. In the ferromagnetic case, where the entries of $J$ are all nonnegative, the classical algorithm by Jerrum and Sinclair \cite{jerrum1993polynomial} gives a poly-time sampler (albeit with a relatively large running-time polynomial), see also \cite{GuoJerrum,FENGGUO}. In the antiferromagnetic case, the problem is more interesting for bounded-degree graphs, where in the case of uniform weights the existence of polynomial-time algorithms is connected to the uniqueness threshold, see \cite{sly2010computational,sly2012computational,galanis2016inapproximability, sinclair2014approximation,li2013correlation}. 

Recently, the development of spectral independence \cite{ALO,Lau} has given tight results on the performance of Glauber dynamics.  This has lead to nearly linear-time algorithms in various settings, see e.g., \cite{chen,chen2022localization,eldan2022spectral,anari2021entropic,koehler2022sampling} and has made it possible to connect the performance of Glauber dynamics with the eigenvalues of the underlying matrix $J$. In this direction,
\cite{eldan2022spectral,anari2021entropic} show that Glauber dynamics is fast mixing when $\lambda_{\max}(J) - \lambda_{\min}(J) < 1$ which  significantly improves upon the standard 
Dobrushin’s uniqueness condition (the latter only applies when $\sum_{j} |J_{ij}| < 1$
for all $i \in [N]$).

On the other side, the optimality of these algorithmic results in terms of the spectrum is less clear. It is known \cite{levin2010glauber, ding2009mixing} that Glauber dynamics mixes slowly in the complete graph for temperatures $\beta>1$, which corresponds precisely to the condition   $\lambda_{\max}(J) - \lambda_{\min}(J) > 1$ 
by taking $J$ to be  the adjacency matrix of the $N$-vertex complete graph, scaled by $\beta/N$. This does not however  translate in a straightforward way to hardness results and does not preclude the possibility that various alternative methods could potentially go beyond the 1-gap, see, e.g., \cite{risteski2016calculate,koehler2022sampling,jain2019mean} for some recent  approaches using variational methods. To this end, Kunisky \cite{kunisky2024optimality} gave further evidence that $\lambda_{\max}(J) - \lambda_{\min}(J) > 1$ is hard for sampling  via a reduction to hypothesis testing in a Wishart
negatively-spiked matrix model that involves random matrices (which is known to resist  low-degree algorithms  \cite{bandeira2020computational}). Kunisky also posed the conjecture that in fact $\NP$-hardness for sampling under spectral constraints should hold when $\lambda_{\max}(J) - \lambda_{\min}(J) > 1$. To add a bit to the mystery, it is noteworthy that the inapproximability results for  the antiferromagnetic case (mentioned earlier) only apply roughly when $\lambda_{\max}(J) - \lambda_{\min}(J) > 2$, see also below for a more detailed discussion.

\paragraph{Our result.}  Our aim in this work is to address Kunisky's conjecture and close the gap between algorithmic and $\NP$-hardness results. In particular, we answer in the affirmative the conjecture in  \cite{kunisky2024optimality}, obtaining $\NP$-hardness results that complement the algorithmic results of \cite{eldan2022spectral,anari2021entropic}. This completes the program initiated in \cite{kunisky2024optimality}, i.e., showing that Glauber is effectively optimal for ``general-purpose'' Ising model sampling, and clarifies the picture in terms of the computational  complexity landscape under spectral constraints.   

To formally state the result, we define the following computational problem.\\

\noindent\fbox{%
    \parbox{\textwidth}{%
\noindent \emph{Problem:} $\SpecIsing(\gamma)$\\
\emph{Input:}
A symmetric matrix $J\in \mathbb{R}^{N\times N}$, with $\lambda_{\max}(J) - \lambda_{\min}(J) < \gamma$.\\
\emph{Output:} The partition function $Z_J=\sum_{\sigma\in\{-1,+1\}^N}\exp\lp(\frac{1}{2}\sigma^\intercal J \sigma\rp)$.
    }%
}

\newcommand{\statetheoremone}{Fix any real $\gamma > 1$. Then, it is $\NP$-hard to approximate \emph{$\SpecIsing(\gamma)$}, even within an exponential factor $2^{cN}$ for some constant $c=c(\gamma)>0$.}
\begin{theorem}
\label{theorem:main}
\statetheoremone
\end{theorem}

This confirms Conjecture 1.9 of  \cite{kunisky2024optimality} and complements the algorithm  of \cite{eldan2022spectral,anari2021entropic}. Using \Cref{theorem:main}, we   get the following result using the standard reduction \cite{JERRUM1986169} from counting to sampling (the problem is self-reducible under scaling of the matrix $J$). Recall, the total variation distance between probability distributions $\mu$ and $\nu$ is defined as $\mathrm{TV}(\mu,\nu) = \frac{1}{2}\|\mu - \nu\|_1$.
\begin{corollary}
For every real $\gamma>1$, the following holds. Suppose there is a poly-time sampler that, on input a symmetric matrix $J\in \mathbb{R}^{N\times N}$ with $\lambda_{\max}(J) - \lambda_{\min}(J) <\gamma$ and $\delta>0$, returns  a configuration $\sigma$ whose distribution is within $\mathrm{TV}$ distance $\delta$ from $\mu_J$. 
Then~$\NP = \RP.$
\end{corollary}

As we will explain next, it is also possible to obtain a more refined version of \Cref{theorem:main}, for the restricted case where each row of the interaction matrix $J$ has at most $d$ non-zero entries, for some fixed integer $d\geq 4$.\vskip 0.1cm

\noindent\fbox{%
    \parbox{\textwidth}{%
\noindent \emph{Problem:} $\BSpecIsing(d,\gamma)$\\
\emph{Input:}
A symmetric matrix $J\in \mathbb{R}^{N\times N}$, with $\leq d$ non-zero entries per row and  $\lambda_{\max}(J) - \lambda_{\min}(J)< \gamma$.\\
\emph{Output:} The partition function $Z_J=\sum_{\sigma\in\{-1,+1\}^N}\exp\lp(\frac{1}{2}\sigma^\intercal J \sigma\rp)$.
    }%
}
\\
\newcommand{\statetheoremmaintwo}{Fix any integer $d\geq 4$ and real $\gamma > \frac{1}{2}\ln(1+\frac{2}{d-3})(d-1+2\sqrt{d-2})$. Then, it is $\NP$-hard to approximate \emph{$\BSpecIsing(d,\gamma)$}, even within an exponential factor $2^{cN}$ for some constant $c=c(\gamma)>0$.}
\begin{theorem}
\label{theorem:main2}
\statetheoremmaintwo
\end{theorem}

Note that when taking the limit $d \to \infty$ in the above bound, we recover the spectral condition $\gamma > 1$ of \Cref{theorem:main}, so asymptotically the bound is tight; we are not aware of algorithmic results that apply specifically to the $d$-sparse setting under the spectral condition. We remark further that applying the results of \cite{sly2012computational,galanis2016inapproximability} would yield hardness only in the setting where $\gamma>d\ln(1+\frac{2}{d-2})$ (see \cite[Section 1.2]{kunisky2024optimality} for a detailed description on how to translate the results), so \Cref{theorem:main2} improves on this by roughly a factor of 2 asymptotically. It should be noted however that the setting in these results is more restrictive (negative weights, which have the same value on all edges) and hence not directly comparable.

\paragraph{Techniques.} Before giving the proofs, we explain briefly the main idea behind \Cref{theorem:main}, the idea for \Cref{theorem:main2} is almost identical, modulo the gadget used in the reduction.

The key ingredient in obtaining \Cref{theorem:main} is to exploit the slow mixing of Glauber dynamics on the complete graph in a suitable way. Recall that \cite{levin2010glauber} showed exponential  mixing time for Glauber dynamics on the $N$-vertex complete graph when the weights on the edges are ferromagnetic equal to $\beta/N$ (entry-wise) for any $\beta>1$ (note that the corresponding matrix $J$ has $\lambda_{\max}(J) - \lambda_{\min}(J)=\beta$). Intuitively, the slow mixing is caused because the distribution exhibits bimodality, i.e., it is  concentrated around two modes/``phases'' corresponding roughly to the all-plus and all-minus configurations (see \Cref{sec:gadget} for more details). Therefore, we would like to use the binary behaviour of the complete graph as a gadget in the reduction.  The main trouble  here is caused by the ferromagnetic interactions which cannot typically be related to NP-hard problems; by contrast, in the antiferrromagnetic case $\beta<0$, the max-probability configurations in the Ising distribution correspond to maximum cuts (when $J$ encodes the adjacency matrix of a graph), and the respective gadgets in the constructions had bipartite structure.\footnote{As a side note, we remark that the factor-2 gap from the antiferromagnetic setting (mentioned below \Cref{theorem:main2}) comes from the use of bipartite gadgets in these results, which have a symmetric spectrum around zero and hence effectively double the range of the eigenvalues.}

Hence, in order to get $\NP$-hardness, we need to introduce some ``mild'' antiferromagnetism (small negative weights): mild to keep the spectrum unchanged and antiferromagnetic to allow us to reduce from an $\NP$-hard problem (we will use MaxCut); this is quite different than the approach of \cite{kunisky2024optimality} where the positive and negative entries in the constructed instance are more heavily mixed up (randomly). At this stage, the main observation is that the previous reductions used in the antiferromagnetic case \cite{sly2010computational,sly2012computational,galanis2016inapproximability} can accommodate this relatively easily; the only difference here is that we need to use small negative weights to connect disjoint copies of the gadgets, and amplify their  effect using appropriately-sized matchings; conveniently, since the matchings (with the small weights on their edges) correspond to a low-rank perturbation, the spectrum of the underlying matrix is close to that of the complete graph. 

The proof of \Cref{theorem:main2} is almost identical. The main difference needed to make our construction sparse is to use a random $d$-regular graph as the gadget, which is known to exhibit slow mixing when $\beta>\beta_d:=\frac{1}{2}\ln(1+\frac{2}{d-2})$ \cite{4389492,dembo2010ising,MosselSly,montanari2012weak}, with a similar bimodal behaviour to that of the complete graph for $\beta>1$. Relative to the spectrum, the well-known result of Friedman~\cite{friedman2008proof} shows that the adjacency matrix $A$ of a random $d$-regular graph satisfies w.h.p. $\lambda_{\max}(A) - \lambda_{\min}(A)\leq \lambda_d+\epsilon$ for any constant $\epsilon>0$, where $\lambda_d:=d+2\sqrt{d-1}$. For technical reasons (see \Cref{remark:optimality} for details),  we need to actually use a $(d-1)$-regular graph as a gadget in the reduction, so the argument sketched above yields $\NP$-hardness when $\gamma>\beta_{d-1} \lambda_{d-1}$ and $d\geq 4$.

\paragraph{Outline and Discussion.} We give the details of the gadget in \Cref{sec:gadget} and the reduction in \Cref{sec:proof}. This gives a self-contained proof of \Cref{theorem:main}; for \Cref{theorem:main2} the argument is identical modulo the use of the (random) $d$-regular graph as the gadget, for which we need to import a couple of non-trivial results from the literature. 

As a final remark before proceeding to the proofs,
 it would be interesting to explore whether the statistical hardness perspective from \cite{kunisky2024optimality} (or some variant) perhaps applies to other counting/sampling problems where $\NP$-hardness results are unlikely, such as approximating the number of independent sets in a bipartite graph \cite{dyer2004relative}, or approximating the partition function of the ferromagnetic Potts model \cite{Potts}. Another related question is  whether such statistical hardness results can be invoked on sparse random graph models where the spectral threshold $\lambda_{\max}(J) - \lambda_{\min}(J)=1$ (that applies to  worst-case instances) is known not to be tight (see \cite{chen2022localization,koehler2022sampling,liu2024fast}).

\section{The Gadget of \Cref{theorem:main}}
\label{sec:gadget}
Our main gadget will be a clique graph $K_n = (V,E)$ with $n$ vertices, where $V=\{1,2,\hdots,n\}$. We will consider $n$ to be an absolute (large) constant that we will choose later. 
For a small integer $t>0$, let $S \subseteq V$ be an arbitrary subset of $V$ with $|S| = t$. Let $r = n - t$. Intuitively, $S$ contains the nodes that will be used to connect the gadgets with each other.  

We define the phase of the configuration $\sigma \in \{-1,1\}^n$ on $V\backslash S$ as 
\[
Y_\sigma =  \mathbf{1}\bigg\{\sum_{i \in V\setminus S}\sigma_i > 0\bigg\}
- \mathbf{1}\bigg\{\sum_{i \in V \setminus S}\sigma_i \leq 0\bigg\}.
\]
Note that the phase of a configuration is defined using only the spins in $V \setminus S$.
For any fixed $\beta > 0$, consider solutions to the equation  
\begin{equation}\label{eq:mean_field}
\ln \frac{1-\alpha}{\alpha}  + 2\beta (2\alpha - 1)=0
\end{equation}
for $\alpha\in [0,1]$. It is not hard to see that  for $\beta>1$ there are exactly three solutions $\alpha=q^-,1/2,q^+$ which satisfy $q^+ - 1/2 = 1/2 - q^->0$. 
Using these, we define the product measure $Q_S^+$ (resp. $Q_S^-)$ on configurations on $S$, where each spin takes the value $+1$ with probability $q^+$, and $-1$ with probability $1-q^+$ (resp. $q^-$ and $1-q^-$). Concretely, for $\tau\in \{-1,+1\}^S$, we have
\begin{equation}\label{eq:product}
Q^\pm_S(\tau) = (q^\pm )^{\frac{\sum_{i \in S}\tau_i + t}{2}} (1 - q^\pm)^{\frac{t - \sum_{i \in S}\tau_i}{2}} = \lp(q^\pm (1-q^\pm)\rp)^{t/2}\lp(\tfrac{q^\pm}{1-q^\pm}\rp)^{\frac{\sum_{i\in S}\tau_i}{2}}\,.
\end{equation}

We now state a lemma that presents the basic properties of the Ising model on our gadget graph. A similar lemma appears in the seminal results of \cite{sly2010computational,sly2012computational}.
Informally, the lemma states that conditioned on the phase of the spins in $V\setminus S$, the spins in $S$ behave almost \emph{independently} from each other, with bias depending on the phase. 

\begin{lemma}\label{lemma:gadget}
Let $\beta>1$. Then, for any real $\epsilon>0$ and integer $t\geq 1$, for all sufficiently large integers $n= n(t,\epsilon)$ such that $n-t$ is odd, the following hold for the Ising model with interaction matrix $J\in \mathbb{R}^{n\times n}$ given by $J = \frac{\beta}{n-t} \mathbf{1}\mathbf{1}^\top$, where $\mathbf{1}$ is the $n$-dimensional vector with all ones.

Let $S\subseteq [n]$ be a subset of the vertices with $|S|=t$. Then:
\begin{enumerate}
    \item \label{item:balance} The  phases on $V\backslash S$ appear with the same probability, i.e., $\Pr_{\sigma\sim \mu_J}[Y_\sigma= +]=\Pr_{\sigma\sim \mu_J}[Y_\sigma= -]=1/2$.

    \item \label{item:product} Conditioned on the phase, the joint distribution of the spins in $S$ is approximately given by the product distribution $Q^{\pm}_S$, i.e., 
    \[\mbox{for any $\tau\in \{-1,+1\}^S$, it holds that  
     $\Pr_{\sigma\sim \mu_J}\big[\sigma_S=\tau \mid Y_\sigma = \pm\big]=(1\pm \epsilon)Q_S^{\pm}(\tau)$.}\]
     \end{enumerate}
\end{lemma}
\begin{proof}
Let $r=n-t$. For $r$ odd (as in the statement of the lemma), we have by symmetry that the phases appear with equal probability. So, we focus on proving the second item. For a vector $x$ with entries $+1$ or $-1$, we denote by $|x|$ the sum of its entries. 

Let $\alpha\in [0,1]$  be such that $\alpha r$ is an integer. For a configuration $\tau\in \{-1,+1\}^S$, let $Z^\alpha(\tau)$ be the contribution to the partition function of configurations $\sigma$ with $\alpha r $ spins from $V\backslash S$ set to $+1$, $(1-\alpha) r $ spins from $V\backslash S$ set to $-1$ and $\sigma_S=\tau$.
Concretely,
\[
Z^\alpha(\tau) = \sum_{\sigma\in \{-1,+1\}^{V};\, \sigma_S=\tau,\,  |\sigma_{V\backslash S}|=(2\alpha-1)r} \exp(\tfrac{1}{2}\sigma^\top J\sigma).
\]
The number of configurations $\sigma$ with $\sigma_S=\tau$ and  exactly $\alpha r$ of the spins in $V\backslash S$ equal to $1$ is ${r \choose \alpha r}$. Using that  $J = \frac{\beta}{r} \mathbf{1}\mathbf{1}^\top$, for each such $\sigma$, we have $\frac{1}{2}\sigma^\top J\sigma=\frac{\beta}{r}(|\sigma_{V\backslash S}|+|\tau|)^2=\frac{\beta}{2r}\big((2\alpha - 1)r+|\tau|\big)^2$. So, 
\begin{equation}\label{eq:Z_def}
Z^{\alpha}(\tau)=\binom{r}{\alpha r}
\exp\lp(\frac{\beta}{2}(2\alpha - 1)^2r + \beta (2\alpha - 1)|\tau| + \frac{\beta}{2r}|\tau|^2\rp).
\end{equation}
We use the well-known approximation of the binomial coefficient using Stirling's approximation. This yields, for any $\alpha \in [0,1]$, that 
\begin{equation}\label{eq:stirling}  {r \choose \alpha r}=\exp(rH(\alpha)+o(r)).
\end{equation}
where $H(\alpha) := -\alpha \ln \alpha - (1-\alpha) \ln (1 - \alpha)$ is the binary entropy function. 
Asymptotically in $r$, we can also ignore the term $\exp(\frac{\beta}{2r}|\tau|^2)$, so we obtain that
\begin{align}\label{eq:Z_upper}
 Z^\alpha(\tau) =\exp\big(r f(\alpha) + o(r)\big) \mbox{ where } f(\alpha) := H(\alpha) + \frac{\beta}{2}(2\alpha - 1)^2.
\end{align}
The function $f(\alpha)$ plays a key role since for large $r$ it controls the asymptotic order of $Z^{\alpha}(\tau)$. The important point, as we will see below, is that the global maximum of $f$ is attained for $\alpha=q^\pm$.

Indeed, we have 
\[f'(\alpha)=-\ln(\alpha)+\ln(1-\alpha)+2\beta(2\alpha-1)\]
and $f''(\alpha)=-\tfrac{1}{\alpha(1-\alpha)}+4\beta$. Since $f''$ has at most two zeros, we have that $f'$ has at most three distinct zeros and hence $f$ has at most three critical points. For $\beta>1$, we have $f'(1/2)=0$ and $f''(1/2)=-4+4\beta>0$, so $f$ has a local minimum at $\alpha=1/2$; therefore, the maximum of $f$ in the interval $[0,1]$ is attained at some point $\alpha\neq 1/2$. Using the symmetry of $f$ around $\alpha=1/2$, there must be at least two global maxima, one in the interval $(0,1/2)$ and $(1/2,1)$. Since $f$ has at most three critical points (and 1/2 is one of them), we conclude that there are exactly two critical points/maxima other than $\alpha=1/2$, which must therefore be the values $q^+,q^-$ as defined in \eqref{eq:mean_field}.

We are now ready to establish the second item of the lemma. We will argue about the $+$ phase, but the other phase is completely symmetric. Let $\tau, \tau' \in \{-1,1\}^S$ be two configurations of spins in $S$. We have that 
\begin{equation}\label{eq:start534}
\frac{\Pr[\sigma_S = \tau |Y(\sigma_{V \setminus S}) = +]}{\Pr[\sigma_S= \tau'|Y(\sigma_{V \setminus S}) = +]} = \frac{\sum_{\alpha > 1/2}Z^\alpha(\tau)}{\sum_{\alpha > 1/2}Z^\alpha(\tau')}.
\end{equation}
We will show that the sums in the numerator and denominator are dominated by $\alpha$ values that are close to $q^+$. First, note that since $q^+$ is the unique global maximum of $f(\alpha)$ in the interval $[1/2,1]$, for any arbitrarily small constant $\delta>0$, there is $\eta>0$ such that $f(\alpha)\leq f(q^+)-3\eta$ for all $\alpha>1/2$ with $\alpha\notin [q^+-\delta,q^++\delta]$. We  pick $\delta>0$ sufficiently small and $r > 0$ sufficiently large so that $\exp(4\beta t\delta+\beta \tfrac{t^2}{r})<\epsilon/2$.  Since $|\tau|\leq t$, it follows that for $r$ large enough it holds that 
\[\sum_{\alpha>1/2;\, |\alpha-q^+|>\delta} Z^\alpha(\tau)\leq \exp (r (f(q^+)-2\eta)).
\]
By the continuity of $f$, for $\alpha=q^++O(1/r)$ we have $f(\alpha)=f(q^+)+O(1/r)$ and therefore 
\[\sum_{\alpha>1/2;\, |\alpha-q^+|\leq \delta} Z^\alpha(\tau)\geq \exp (r (f(q^+)-\eta)).\]
It follows that
\begin{equation}
    \label{eq:bound}
    \frac{\sum_{\alpha>1/2;\, |\alpha - q^+| > \delta}Z^\alpha(\tau)}{\sum_{\alpha>1/2;\, |\alpha - q^+| \leq \delta}Z^\alpha(\tau)} \leq \exp(-\eta r)\leq \epsilon/2.
\end{equation}
for all sufficiently large $r$. 
Thus,
\begin{align}
\frac{\sum_{\alpha > 1/2}Z^\alpha(\tau)}{\sum_{\alpha > 1/2}Z^\alpha(\tau')} 
& \leq
\frac{\sum_{\alpha > 1/2}Z^\alpha(\tau)}{\sum_{|\alpha - q^+| \leq \delta} Z^\alpha(\tau')}
=\frac{\sum_{\alpha > 1/2}Z^\alpha(\tau)}{\sum_{|\alpha - q^+| \leq \delta} Z^\alpha(\tau)}\cdot \frac{\sum_{|\alpha - q^+| \leq \delta} Z^\alpha(\tau)}{\sum_{|\alpha - q^+| \leq \delta} Z^\alpha(\tau')}\nonumber
\\
& \leq
(1+\epsilon/2)\frac{\sum_{|\alpha - q^+| \leq \delta}Z^\alpha(\tau)}{\sum_{|\alpha - q^+| \leq \delta}Z^\alpha(\tau')}\label{eq:fer3442},
\end{align}
where the last inequality follows from \eqref{eq:bound}. 

On the other hand, for any $\alpha$ with $|\alpha - q^+| \leq \delta$, 
using \eqref{eq:Z_def} we get
\begin{align}\label{eq:ratio_upper}
\frac{Z^\alpha(\tau)}{Z^\alpha(\tau')} &= \exp\lp(\beta(2\alpha - 1)(|\tau| - |\tau'|) + \tfrac{\beta(|\tau|^2 - |\tau'|^2)}{2r}\rp)\leq \exp(4\beta t\delta+\beta \tfrac{t^2}{r}) \exp\lp(\beta(2q^+ - 1)(|\tau| - |\tau'|)\rp)\nonumber\\ &\leq (1+\epsilon/2) \exp\lp(\beta(2q^+ - 1)(|\tau| - |\tau'|)\rp), 
\end{align}
where the last inequality follows from the choice of $\delta$ and $r$. Using the definition \eqref{eq:product} and the fact that $q^+$ is a solution of \eqref{eq:mean_field}, i.e., that $f'(q^+)=0$, we have that
\[\exp\lp(\beta(2q^+ - 1)(|\tau| - |\tau'|)\rp)=  \lp(\frac{q^+}{1-q^+}\rp)^{\frac{|\tau|-|\tau'|}{2}} =\frac{Q_S^+(\tau)}{Q_S^+(\tau')}.\]
Hence, from \eqref{eq:ratio_upper} we obtain that $\frac{Z^\alpha(\tau)}{Z^\alpha(\tau')} \leq (1+\epsilon/2)\frac{Q_S^+(\tau)}{Q_S^+(\tau')}$. Since this holds for all $\alpha$ with $|\alpha - q^+| \leq \delta$, we have
\begin{equation}\label{eq:two_ratios}
 \frac{\sum_{|\alpha - q^+| \leq \delta}Z^\alpha(\tau)}{\sum_{|\alpha - q^+| \leq \delta}Z^\alpha(\tau')}\leq (1+\epsilon/2)\frac{Q_S^+(\tau)}{Q_S^+(\tau')}.
\end{equation} 
Combining this with \eqref{eq:start534} and \eqref{eq:fer3442}, we obtain that 
\[\frac{\Pr[\sigma_S = \tau |Y(\sigma_{V \setminus S}) = +]}{\Pr[\sigma_S= \tau'|Y(\sigma_{V \setminus S}) = +]}\leq (1+\epsilon)\frac{Q_S^+(\tau)}{Q_S^+(\tau')}.\] 
By interchanging the roles of $\tau,\tau'$, we also obtain the inverse inequality, so
\begin{equation}\label{eq:bound343454}
    (1-\epsilon)\frac{Q_S^+(\tau)}{Q_S^+(\tau')} \leq \frac{\Pr[\sigma_S = \tau |Y(\sigma_{V \setminus S}) = +]}{\Pr[\sigma_S= \tau'|Y(\sigma_{V \setminus S}) = +]}\leq (1+\epsilon)\frac{Q_S^+(\tau)}{Q_S^+(\tau')}.
\end{equation}
For $\tau \in \{-1,1\}^S$, observe that we can  expand the ratio
\begin{align*}
    \frac{\Pr[\sigma_S=\tau | Y = +]}{Q_S^{+}(\tau)} =  \frac{\sum_{\tau'} Q^+_S(\tau')\Pr[\sigma_S=\tau \mid Y = +]}{\sum_{\tau'} Q_S^{+}(\tau)\Pr[\sigma_S=\tau'\mid  Y = +]}
\end{align*}
so using that $\min_i \frac{a_i}{b_i}\leq \frac{\sum_i a_i}{\sum_i b_i}\leq \max_i \frac{a_i}{b_i}$ for non-negative $(a_i)_i, (b_i)_i$, we obtain from \eqref{eq:bound343454} that
\[\left |
    \frac{\Pr[\sigma_S=\tau | Y = +]}{Q_S^{+}(\tau)} - 1\right|\leq \max_{\tau'}\left| \frac{Q^+_S(\tau')\Pr[\sigma_S=\tau \mid Y = +]}{ Q_S^{+}(\tau)\Pr[\sigma_S=\tau'\mid  Y = +]}-1\right|\leq \epsilon. \]
    This finishes the proof.
    \end{proof}

\section{Proofs of Main Results}
\subsection{Proof of \Cref{theorem:main}}
\label{sec:proof}
Let $\gamma > 1$ and $\beta=(1+\gamma)/2>1$. Following the technique in \cite{sly2010computational,sly2012computational, galanis2016inapproximability}, we reduce \textsc{MaxCut} on 3-regular graphs to $\SpecIsing(\gamma)$. 

Consider a 3-regular graph $H = (V_H,E_H)$ with $|V_H|=m$ vertices, an instance of \textsc{MaxCut}. Let $G$ be the clique graph on $n$ vertices, with a subset $S$ of the vertices with $|S|=t$ that will be used as terminals (cf. \Cref{lemma:gadget}); for convenience, we assume that $t>0$ is a multiple of 3 (with $n\gg 3t$). We construct an instance $H^G$ of $\SpecIsing(\gamma)$  as follows:
\begin{itemize}
    \item We replace each node $v \in V_{H}$ with a distinct copy of  the gadget clique graph $G$. In particular, for any $v \in V_H$, consider a copy $G_v = (W_v, E_v)$ of the gadget $G$; each edge in $E_v$ has weight $w_+=\beta/r$ as in \Cref{lemma:gadget}, where recall that $\beta = (1 + \gamma)/2 > 1$. For each $v \in V_H$, let $S_v \subseteq W_v$ be a subset of the vertices in $G_v$ of size $t = n  - r$. 
    Let $\wh{H}^G$ be the disjoint union of the $G_v$'s for $v \in H$.
    Note that the number of vertices of $\wh{H}^G$ is $n m$. 
    
    \item We now describe how to encode the edges of $H$ using connections between the gadgets (which will complete the construction of $H^G$). Assume that the node $u\in V_H$ has neighbors $v_1,v_2,v_3$ in $H$, i.e., $(u,v_i) \in E_H, i = 1,\ldots,3$. Then, we partition $S_u$ into subsets $S_u^i$ of size $t/3$ each. Each subset $S_u^i$ corresponds to one of the three neighbors of $u$. Then, for each $i = 1,2,3$, we add a perfect matching between $S_u^i$ and the corresponding subset $S_{v_i}^j$ of $S_{v_i}$ that corresponds to $u$.
    The weight of each of these edges in the matching will be $w_-=(1-\gamma)/5 < 0$, since $\gamma > 1$. 
    This antiferromagnetic structure across different copies will be crucial in order to approximate $\mathsf{maxcut}(H)$ by approximating the partition function of $H^G$.
\end{itemize}

Let $J$ be the adjacency matrix of the weighted graph $H^G$. We first show that the spectrum of $J$ has the desired properties, i.e., that $\lambda_{\max}(J) - \lambda_{\min}(J) < \gamma$. 

\begin{claim}
[Structure of $H^G$]
\label{claim:1}
The symmetric matrix $J = D + E \in \mathbb{R}^{nm  \times nm}$, where $D$ is a block diagonal matrix where the matrix of each block of size $n \times n$ is $\frac{\beta}{r}\mathbf{1}\mathbf{1}^\top$ and $E$ contains in each row exactly one non-zero element of magnitude $(1-\gamma)/5$.
\end{claim}
\begin{proof}
    By construction since $n$ is the number of vertices of the gadget and $m$ is the number of vertices of the input graph.
\end{proof}

\begin{claim}
[Spectrum Preservation]
\label{lemma:spectrum}
For any integer $t > 0$, there exists $n(t,\gamma) >0$, such that for $n > n(t,\gamma)$ it holds $|\lambda_{\max}(J) - \lambda_{\min}(J)| < \gamma$. 
\end{claim}
\begin{proof}
We will use \Cref{claim:1}.
Using Weyl's inequality (see Chapter 3 in \cite{bhatia2007perturbation}), which controls the eigenspectrum of a matrix under small perturbations in each entries, we have that for any $i$, it holds that 
$|\lambda_i(J) - \lambda_i(D)| \leq \|E\|,$
where $\|E\|$ is the spectral norm of $E$. 
By definition, $E$ has one element in each row of absolute value $(\gamma - 1)/5$, so $
\|E\| \leq \frac{\gamma -1}{5}.
$
It follows that
\begin{align}\label{eq:spectrum}
|\lambda_{\max}(J) - \lambda_{\min}(J)| &\leq
|\lambda_{\max}(J) - \lambda_{\max}(D)| + |\lambda_{\max}(D) - \lambda_{\min}(D)| +
|\lambda_{\min}(D) - \lambda_{\min}(J)| \nonumber \\
&\leq 
\frac{2(\gamma -1)}{5} + \frac{n}{r} \frac{1 + \gamma }{2}.
\end{align}
In the above we used the well-known fact that the spectrum of $D$ is the spectrum of each of the blocks, which, in turn, is equal to
\[
\lambda_{\max}(D) - \lambda_{\min}(D) = \frac{n}{r} \frac{1 + \gamma}{2},
\]
since each block is a rank-$1$ matrix. Now, recall that $n = r + t$, so by choosing $r$ sufficiently large we can make $\frac{n}{r} < \frac{6\gamma + 4}{5 \gamma + 5}$, which implies that the right hand side in \eqref{eq:spectrum} is $< \gamma$. 
\end{proof}

We next show that
if we could approximate $Z_J$ within an arbitrarily small exponential factor in poly-time, we would obtain a PTAS for $\mathsf{maxcut}(H)$.  This part of the argument is largely based on the techniques of \cite{sly2012computational}; we first state the following lemma whose proof is given for completeness in \Cref{appendix:proof}.

\begin{lemma}\label{lemma:estimate}
It holds that
\[
(1-4\eps)^m 2^{-m}
\leq 
\frac{Z_{H^G}/Z_{\wh{H}^G}}{A^{3mt/2}\lp(B/A\rp)^{\mathsf{maxcut}(H)t/3}}
\leq (1 +4\eps)^m,
\]
where $A,B$ are positive constants depending only on $\gamma$ (and are explicitly defined in \eqref{eq:AB}).
\label{lemma:approximation}
\end{lemma}

With these pieces at hand, we are now ready to complete the reduction for Theorem~\ref{theorem:main}, which we restate here for convenience.
\begin{theoremone}
\emph{\statetheoremone}
\end{theoremone}
\begin{proof}
Assume  that for any arbitrarily small   constant $\delta> 0$, there is an oracle $\mathsf{approx}_\delta$ such that, for any $J$
with $\lambda_{\max}(J) - \lambda_{\min}(J) \leq \gamma$, we have that, when $F = \mathsf{approx}_\delta(J)$,
$|F - \log(Z(J))| \leq \delta m$. We will show how to obtain a PTAS for $\textsc{MaxCut}$ on 3-regular graphs, i.e., approximate $\textsc{MaxCut}$ on 3-regular graphs within an arbitrarily small factor.

Let $H$ be a 3-regular graph $H$ on $m$ vertices, an instance of $\textsc{MaxCut}$. The maximum cut of $H$ is at least  the expected value of a random cut which is equal to $3m/4$. We then construct $H^G$ and $\wh{H}^G$ as above. Observe that $Z(\wh{H}^G)$ can be computed in poly-time since $\wh{H}^G$ is a disjoint collection  of constant-size gadget graphs.  Moreover, by \Cref{lemma:spectrum}, $H^G$ is an instance of $\SpecIsing(\gamma)$. So, 
 we can use the oracle $\mathsf{approx}_\delta$ on $H^G$, which will give us an output $F_H$ with 
 the guarantee
 \[
 |F_H - \log Z_{H^G}| \leq \delta m n.
 \]
 \Cref{lemma:approximation} implies that
 \[
 \frac{3\log \lp(\frac{Z_{H^G}/Z_{\wh{H}^G}}{A^{3mt/2}(1+4\eps)^m}\rp)}{t\log (B/A)}
 \leq 
 \mathsf{maxcut}(H)
 \leq 
\frac{3\log \lp(\frac{2^mZ_{H^G}/Z_{\wh{H}^G}}{A^{3mt/2}(1-4\eps)^m}\rp)}{t\log (B/A)}.
 \]
Thus, by using the output $F_H$ we can compute upper and lower bounds for the maximum cut value, which differ by $O((\delta n + 1 )m/t)$. 
Since $m \leq 4/3 \mathsf{maxcut}(H)$, to show the desired PTAS for \textsc{MaxCut}, it only remains to show that the quantity $R=(\delta n + 1)/t$ can be made arbitrarily small, say less than some  target value $\zeta$, where $\zeta>0$ is an arbitrary constant. We first take $t$ to be sufficiently large, so that $1/t<\zeta/2$ is sufficiently small. This makes $n$ to be large, but still a constant, and hence $n/t$ is a constant.  So, by taking $\delta$ small enough, we will have $\delta n /t<\zeta/2$, making $R<\zeta$ as desired.

This yields  the desired PTAS. Since \textsc{maxcut} is $\APX$-hard \cite{alimonti1997hardness}, we conclude that it is $\NP$-hard to approximate $Z_J$ within some exponential factor, as wanted.
\end{proof}
\subsection{Proof of 
\Cref{theorem:main2}}
For integers $d,n\geq 3$ with $dn$ even, let $G_{n,d}$ be a $d$-regular graph chosen uniformly at random among all such graphs with vertex set $V=\{1,2,\hdots,n\}$.  Let $S\subseteq [n]$ be an arbitrary subset of the vertices of size $t$. Consider the Ising distribution $\mu_J$ with $J=\beta A$ where $A$ is the adjacency matrix of $G$ and $\beta>\frac{1}{2}\ln(1+\frac{2}{d-2})$. 

The range of $\beta$ corresponds to the so-called non-uniqueness regime on the $d$-regular tree; roughly, this implies that on the $d$-regular tree of height $h$, when we condition the leaves to be $+$ and take the limit $h\rightarrow \infty$, the marginal probability that the root is plus converges to some value $q^+>1/2$. Similarly, when we condition the leaves to be $-$, the marginal probability that the root is plus converges to some value $q^-<1/2$.\footnote{\label{fn:qpqm}To define $q^+,q^-$ more explicitly, for $\beta>\frac{1}{2}\ln(1+\frac{2}{d-2})$, let $\tilde{q}^+>1>\tilde{q}^->0$ be the solutions of
$x=\Big(\frac{\exp(2\beta) x+1}{x+\exp(2\beta)}\Big)^{d-1}$. Then,  
$q^+,q^-$ are defined from $\frac{q^+}{1-q^+}=\tilde{q}^+\frac{\exp(2\beta)\tilde{q}^+ +1}{\tilde{q}^++\exp(2\beta)}$ and $\frac{q^-}{1-q^-}=\tilde{q}^-\frac{\exp(2\beta)\tilde{q}^- +1}{\tilde{q}^-+\exp(2\beta)}$, see also \cite[Section 3]{galanis2016inapproximability}.}

It is well-known by now \cite{dembo2010ising, montanari2012weak} that this behaviour on the tree manifests itself on the random $d$-regular graph, roughly because of the tree-like neighborhoods in the latter. To make this more precise in our setting,  analogously to \Cref{sec:gadget}, for a subset $S\subseteq V$, define the phase $Y_S(\sigma)$ of a configuration $\sigma\in \{-1,+1\}^V$ to be $+$ if $\sum_{i\in V\backslash S} \sigma_i\geq 0$, and $-$ otherwise.
We also define the product measures $Q_S^\pm$ on $S$ analogously to \eqref{eq:product}, using now the values of $q^+,q^-$ as defined above (see also \Cref{fn:qpqm}). 
Then, the following lemma captures the main properties of the gadget that we need.
\newcommand{\statelemgadgettwo}{Let $d\geq 3$ be an integer and $\beta>\frac{1}{2}\ln(1+\frac{2}{d-2})$. Then, for any real $\epsilon>0$ and integer $t\geq 1$, for all sufficiently large integers $n= n(t,\epsilon)$ with $n-t$ odd, the following holds with probability $1-\epsilon$ over the choice of $G\sim G_{n,d}$. Let $S\subseteq V$ be a subset of vertices with $|S|=t$.

Consider the Ising model with interaction matrix $J=\beta A$ where $A$ is the adjacency matrix of $G$.  
Then:
\begin{enumerate}
\item $\lambda_{\max}(J) - \lambda_{\min}(J)\leq \beta(d+2\sqrt{d-1})+\epsilon$.

    \item The  phases appear with the same probability, i.e., $\Pr_{\sigma\sim \mu_J}[Y_\sigma= +]=\Pr_{\sigma\sim \mu_J}[Y_\sigma= -]=1/2$.

    \item Conditioned on the phase, the joint distribution of the spins in $S$ is approximately given by the product distribution $Q^{\pm}_S$, i.e., 
    \[\mbox{for any $\tau\in \{-1,+1\}^S$, it holds that  
     $\Pr_{\sigma\sim \mu_J}\big[\sigma_S=\tau \mid Y_\sigma = \pm\big]=(1\pm \epsilon)Q_S^{\pm}(\tau)$.}\]
     \end{enumerate}
     }
\begin{lemma}
\label{lemma:gadget2}
\statelemgadgettwo
\end{lemma}
\begin{proof}
The first item is Friedman's result \cite{friedman2008proof}, see also \cite{bordenave2020new}. The second item is by symmetry of the configuration space (since $n$ is odd). The third item follows by \cite[Theorem 2.4]{montanari2012weak}, see also \cite[Theorem 2.7]{dembo2010ising} and \cite[Proposition 4.2]{sly2012computational} for related results. Technically, there is a bit of work to translate the results here, we give the details for the interested reader in \cref{sec:lemgadgetwo}.
\end{proof}
\begin{remark}\label{rem:det}
We will use the gadget of \Cref{lemma:gadget2} for some large but otherwise constant value of $n$. So, we can find a $d$-regular graph $G$ satisfying Items 1-3 of \Cref{lemma:gadget2} in deterministic time. 
\end{remark}
We are now ready to prove \Cref{theorem:main2}, which we restate here for convenience.

\begin{theoremtwo}
\emph{\statetheoremmaintwo}
\end{theoremtwo}
\begin{proof}
Let  
\begin{equation}\label{eq:betadlambdad}
\beta_{d-1} := \tfrac{1}{2} \ln \big(1 + \tfrac{2}{d-3}\big), \quad \lambda_{d-1} := 
d-1 + 2\sqrt{d-2}
\end{equation}
and set $\beta=\beta_{d-1}+\eta$, $\lambda=\lambda_{d-1}+\eta$ where $\eta>0$ is a small constant so that $\beta \lambda+2\eta<\gamma$ (note that such an $\eta$ exists since $\gamma>\beta_{d-1}\lambda_{d-1}$).

Assume that we are given a 3-regular instance $H$ of \textsc{MaxCut} with $m$ vertices. Let $G$ be a $(d-1)$-regular gadget with $n$ vertices for some sufficiently large $n$, i.e., $G$ satisfies Items 1-3 of \Cref{lemma:gadget2} for degree $d-1$ and $\beta=\beta_{d-1}+\eta$, see also \Cref{rem:det}. So, according to Item~1 there,  the interaction matrix $J_G$ corresponding to $G$ satisfies $\lambda_{\max}(J_G)-\lambda_{\min}(J_G) \leq \beta \lambda$.

Using $G$, the construction of the graph $H^G$ is identical to that of \Cref{sec:proof}, i.e., we have a distinct copy of $G$ for each node of $H$ and,  for each pair of neighbouring nodes of $H$, we add a matching of size $t/3$ between the corresponding gadgets using the vertices in $S$. Note that $H^G$ has maximum degree $d$, so the interaction matrix of $H^G$, denoted by $J$ henceforth, has at most $d$ non-zero entries per row.

The weight of an edge inside the gadget is $w_+ = \beta > 0$  and the weight of the edges that connect two gadgets is $w_{-} = -\eta< 0$ (antiferromagnetic connections).
Analogously to \Cref{claim:1}, the symmetric matrix $J$ can be written as $ D + E \in \mathbb{R}^{nm  \times nm}$, where (i) $D$ is a block diagonal matrix with the matrix in each block being the $n\times n$ adjacency matrix of $G$  scaled by 
$w_+$, and (ii) $E$ contains in each row exactly one non-zero element of magnitude $w_-$. The same argument as in the proof of \Cref{lemma:spectrum} gives that
\begin{align*}
|\lambda_{\max}(J) - \lambda_{\min}(J)|&\leq 
|\lambda_{\max}(J) - \lambda_{\max}(D)| + |\lambda_{\max}(D) - \lambda_{\min}(D)| +
|\lambda_{\min}(D) - \lambda_{\min}(J)|\\ 
&\leq 2 \eta + \beta \lambda < \gamma.
\end{align*}
This establishes that $H^G$ is a valid instance of $\BSpecIsing(d,\gamma)$.

Now, using Item 3 of \Cref{lemma:gadget2}, we obtain the exact same  estimate as in \Cref{lemma:estimate} (with the same expressions for the constants $A,B$ modulo the new values of $w_+$ and $w_-$), and therefore the same argument used in the proof of \Cref{theorem:main} applies verbatim to show $\NP$-hardness of approximating the partition function within an arbitrarily small exponential factor.
\end{proof}
\begin{remark}\label{remark:optimality}
Note that we could make the graph $H^G$ to be $d$-regular for any integer $d\geq 3$ by taking the gadget $G$ to be a random $d$-regular graph with a matching of size $t$ removed (and using the endpoints of the matching as the set $S$ of terminals); this more refined construction has been used for example in the hardness results of \cite{sly2010computational,sly2012computational,galanis2016inapproximability}. While one can show the analogue of Items 2 and 3 with minor modifications (analogously to what was done in the proof of \Cref{lemma:gadget2}), the proof of Item 1 for this modified gadget seems to require more careful adaptation of the proofs in \cite{friedman2008proof,bordenave2020new}. It is nevertheless reasonable to expect that the same bound on the range of the eigenvalues as stated currently in Item 1 will still apply; provided this is indeed the case, one can improve slightly the parameters of \Cref{theorem:main2} to $d\geq 3$ and $\gamma>\beta_d\lambda_d$, where $\beta_d,\lambda_d$ are as in \eqref{eq:betadlambdad}.
\end{remark}

\bibliography{bib}

\appendix
\section{Proof of \Cref{lemma:approximation}}
\label{appendix:proof}
\begin{proof}
[Proof of \Cref{lemma:approximation}]
We follow the same proof approach as in Lemma 4.3 in \cite{sly2012computational}.
For $v \in H$ and a configuration $\sigma$ on $H^G$, define random variable $Y_v = Y_v(\sigma)$ to be the phase of the gadget $G_v$ under $\sigma$, as discussed in \Cref{sec:gadget}. Let $\calY = \calY(\sigma) = \{Y_v\}_{v \in H} \in \{-1,+1\}^m$ be the vector of phases of all the gadgets, for a particular configuration $\sigma$ of the spins. For any fixed vector of phases $\calY' \in \{-1,+1\}^m$, 
denote
\begin{align*}
Z_{H_G}(\calY') := \sum_{\sigma \in \{1,-1\}^{mn}}\exp\lp(\frac{1}{2}\sigma^\top J \sigma\rp) \mathbbm{1}\{\calY(\sigma) = \calY'\}\,,\\
Z_{\wh{H}_G}(\calY') := \sum_{\sigma \in \{1,-1\}^{mn}}\exp\lp(\frac{1}{2}\sigma^\top D \sigma\rp) \mathbbm{1}\{\calY(\sigma) = \calY'\}\,.
\end{align*}
Using the first Item of \Cref{lemma:gadget},
we have that
\begin{equation}\label{eq:ratio_hat}
\lp(\frac{1}{2} - \epsilon\rp)^m
\leq
\frac{Z(\wh{H}^G; \calY)}{Z(\wh{H}^G)} \leq \lp(\frac{1}{2} + \eps\rp)^m
\end{equation}
for all phases $\calY\in \{-1,+1\}^m$.
For simplicity, define the function 
\[
\psi(x,y) := \exp\lp(2w_-xy\rp) \mbox{ where  recall that $w_-=(1-\gamma)/5$}.
\]
Observe that we can write the ratio 
\[
\frac{Z_{H^G}( \calY)}{Z_{\wh{H}^G}(\calY)}
=
\sum_{\sigma \in \{-1,1\}^{mn}} \prod_{v \in H}
\Pr_{G_v}[\sigma_{W_v} | Y_v]
\prod_{(i,j) \in E(H^G) \setminus E(\wh{H}^G)} \psi(\sigma_i,\sigma_j)\,.
\]
Now, by \Cref{lemma:gadget}, the right hand side
is within a $(1 \pm \eps)^m$ factor of
\[
\sum_{\sigma \in \{-1,1\}^{tm}} \prod_{v \in H}
Q^{Y_v}[\sigma_{S_v} ]
\prod_{(i,j) \in E(H^G) \setminus E(\wh{H}^G)} \psi(\sigma_i,\sigma_j)
\]
which can be calculated exactly since it is a product distribution. For $i,j \in V(H^G)$, let $\E_{ij}[\cdot|\calY]$ denote the expectation with respect to the distribution where $\sigma_i,\sigma_j$ are independent and each marginal is given by either $Q^+$ or $Q^-$, depending on the phase vector $\calY$.
\begin{align*}
\sum_{\sigma \in \{-1,1\}^{tm}} \prod_{v \in H}
Q^{Y_v}[\sigma_{S_v} ]
\prod_{(i,j) \in E(H^G) \setminus E(\wh{H}^G)} \psi(\sigma_i,\sigma_j) &= 
\prod_{(i,j) \in E(H^G) \setminus E(\wh{H}^G)} \E_{i,j}[\psi(\sigma_i,\sigma_j)|\calY]\,.
\end{align*}
For $i \in G_u$ and $j \in G_v$, we have that
\[
\E_{i,j}[\psi(\sigma_i,\sigma_j)|\calY] = \left\{
\begin{array}{ll}
      A, \quad\text{ if $Y_u = Y_v$}\\
      B, \quad\text{ if $Y_u \neq Y_v$}
\end{array} 
\right.
\]
where
\begin{equation}\label{eq:AB}
\begin{aligned}
A:=  \lp((q^+)^2 + (1-q^+)^2\rp)\psi(1,1) + 2q^+(1-q^+)\psi(1,-1),\\
B: = \lp((q^+)^2 + (1-q^+)^2\rp)\psi(-1,1) + 2q^+(1-q^+)\psi(1,1).
\end{aligned}
\end{equation}
Since $\gamma > 1$, we have $\psi(1,1) < 1 < \psi(-1,1)$, thus $B > A$. 
Now, by the construction/structure of $H^G$ in Section~\ref{sec:proof}, we have that
\[
\prod_{(i,j) \in E(H^G) \setminus E(\wh{H}^G)} \E_{i,j}[\psi(\sigma_i,\sigma_j)|\calY] =
A^{3mt/2}\lp(\frac{B}{A}\rp)^{\mathsf{cut}(\calY)t/3}\,.
\]
Now, together with \eqref{eq:ratio_hat} this implies
that
\begin{align*}
Z_{H^G} = \sum_{\calY \in \{-1,1\}^m} \frac{Z_{H^G}( \calY)}{Z_{\wh{H}^G}(\calY)} Z_{\wh{H}^G}(\calY) 
\leq 2^m (1 + \eps)^m \lp(\frac{1}{2} + \eps\rp)^m A^{3mt/2} \lp(\frac{B}{A}\rp)^{\mathsf{maxcut}(H) t/3} Z_{\wh{H}^G}
\end{align*}
and 
\begin{align*}
Z_{H^G} = \sum_{\calY \in \{-1,1\}^m} \frac{Z_{H^G}( \calY)}{Z_{\wh{H}^G}(\calY)} Z_{\wh{H}^G}(\calY) 
\geq  (1 - \eps)^m \lp(\frac{1}{2} - \eps\rp)^m A^{3mt/2} \lp(\frac{B}{A}\rp)^{\mathsf{maxcut}(H) t/3} Z_{\wh{H}^G}\,.
\end{align*}
Rearranging gives the required result. 
\end{proof}

\section{Proof of \Cref{lemma:gadget2}}\label{sec:lemgadgetwo}
Here we give the details of how to deduce Lemma~\ref{lemma:gadget2} from the results of \cite{montanari2012weak}. We restate the lemma here for convenience. 
\begin{lemgadgetwo}
\emph{\statelemgadgettwo}
\end{lemgadgetwo}
\begin{proof}
The first item is Friedman's result \cite{friedman2008proof}, see also \cite{bordenave2020new}. The second item is by symmetry of the configuration space (since $n$ is odd). 
Let's focus on proving the third item. Let $\mathbb{T}_d$ denote the rooted infinite $d$-regular tree and $\mathbb{T}^t_d$ be the finite subtree of $\mathbb{T}_d$ that contains all nodes at distance at most $t$ from the root. Let $\nu^+_t$ (resp. $\nu^-_t$) be the distribution of an Ising model on $\mathbb{T}_d^t$ with parameter $\beta$ where all nodes at distance $t$ from the root are conditioned to be $+1$ (resp. $-1$). Then we can define the Gibbs measure $\nu^+$ (resp. $\nu^-$) on $\mathbb{T}_d$ to be the weak limit of $\nu^+_t$ (resp. $\nu^-_t$) as $t \to \infty$. For more details see \cite{georgii2011gibbs}.

We will use \cite[Theorem 2.4]{montanari2012weak}; it is a standard fact that $G_{n,d}$ satisfies w.h.p. the assumptions of the theorem, i.e.,  $G_{n,d}$ is locally tree-like and an expander. Therefore, assume that we have a  Finally, let $\mu_n$ be the Ising model measure as defined in the statement of the Lemma and $\mu_n^+$ (resp. $ \mu_n^-$) be the conditional distribution of $\mu_n$, conditioned on $\sum_{i \in V} \sigma_i \geq 0$ (resp. $\sum_{i \in V} \sigma_i < 0$). 
We know that $G_{n,d}$ satisfies the assumptions of \cite[Theorem 2.4]{montanari2012weak} with probability $1 - o(1)$, so all the subsequent statements hold with high probability. Then, by \cite[Theorem 2.4]{montanari2012weak} (see also \cite[Theorem 2.7]{dembo2010ising}), we know that $\mu_n^+$ (resp. $\mu_n^-$) converges locally in probability to $\nu^+$ (resp. $\nu^-$). 
Moreover, by \cite[Theorem 2.5]{montanari2012weak} (applied to the function $f_{i,n}(\sigma) = \sigma_i$ for $\sigma\in \{-1,+1\}^n$) where $\sigma \sim \mu_n^+$
\begin{equation*}
   \lp| \frac{1}{n} \sum_{i=1}^n \sigma_i - \E_{\mu_n^+}\lp[\frac{1}{n} \sum_{i=1}^n \sigma_i\rp]\rp| \to 0
\end{equation*}
in probability. Also, by \cite[Theorem 2.4]{montanari2012weak} $\mu_n^+$ converges locally in probability to $\nu^+$, which implies that
\begin{equation*}
    \lp| \E_{\nu^+}[\sigma_o]-\E_{\mu_n^+}\lp[\frac{1}{n} \sum_{i=1}^n \sigma_i\rp]\rp| \to 0
\end{equation*}
where $\sigma_o$ is the spin of the root of $\mathbb{T}_d$. We conclude that, under $\mu_n^+$
\[
\frac{1}{n} \sum_{i=1}^n \sigma_i \to \E_{\nu^+}[\sigma_o]
\]
in probability. Similarly, under $\mu_n^-$
\[
\frac{1}{n} \sum_{i=1}^n \sigma_i \to \E_{\nu^-}[\sigma_o]
\]
in probability. 
These imply that under $\mu_n = \frac{1}{2}(\nu^- + \nu^-)$
\begin{equation}\label{eq:phase_lim}
    \frac{1}{n}\mathcal{Y}(\sigma) \sum_{i \in V} \sigma_i \to \frac{1}{2}\lp(\E_{\nu^+}[\sigma_0] - \E_{\nu^-}[\sigma_0]\rp) = \nu^+(\sigma_o = 1) - \nu^-(\sigma_o = 1)
\end{equation}
in probability. 
We now use an argument similar to \cite[Proposition 4.2]{sly2012computational}. 
For any integer $l > 0$, let $B_l$ be the union over all $u \in S$ of the $l$-hop neighborhood of $u$. For any fixed $l > 0$, with probability $1-o(1)$ these neighborhoods will be disjoint from each other. Let $R_l$ be the nodes in the boundary of $B_l$. Let $\mathcal{Y}_l(\sigma) = \sum_{i \in V \setminus B_l} \sigma_i$ be the phase defined only using the spins outside of $B_l$. By \cite[Theorem 2.4]{montanari2012weak}, we know that $\mu_n^+(\sigma_{B_t} = \cdot)$ converges to $\nu^+(\sigma_{B_t} = \cdot)$. 
Using \eqref{eq:phase_lim} we conclude that for any $l > 0$, with probability $1 - o(1)$ we have $\mathcal{Y}(\sigma) = \mathcal{Y}_l(\sigma)$. Thus, \cite[Theorem 2.4]{montanari2012weak} implies that $\mu_n(\sigma_{B_t} = \cdot | \mathcal{Y}_l = +)$ converges to $\nu^+(\sigma_{B_t} = \cdot)$ as well.
For any $\epsilon > 0$, define the set
\[
U_\eps^l(u) := \{\eta_l \in \{-1,1\}^{|B_l|}: |\mu_n(\sigma_u = + |  \sigma_{B_l} = \eta_l) - \nu^+(\sigma_u = +)| > \eps\}.
\]

Intuitively, our goal is to show that $U_\eps$ will have small mass under $\mu_n(\cdot| \mathcal{Y}_l = +)$ for sufficiently large $l$.
Formally, we would like to prove that
\begin{equation}\label{eq:limit}
    \limsup_{l \to \infty} \lim_{n \to \infty}\mu_n(U_\eps^l(u)|\mathcal{Y}_l=+) = 0.
\end{equation}
Suppose we have \eqref{eq:limit}. Then, 
by a union bound over all $u \in S$ (since the size of the set $|S| =t$ is considered fixed) we obtain
\[
\limsup_{l \to \infty} \lim_{n \to \infty}\mu_n(\cup_{u \in S} U_\eps^l(u)|\mathcal{Y}_l=+) = 0.
\]
This implies that for any $\eps>0$, we can find $n = n(\eps) ,l = l(\eps)$ such that
\[
\Pr_{\eta}\lp( \forall u \in S:|\mu_n(\sigma_u = + |  \sigma_{B_l} = \eta) - \nu^+(\sigma_u = +)| < \eps
|\mathcal{Y}_l = +\rp) > 1 - \eps.
\]
Hence,  by law of total probability and the conditional independence of the spins in $S$ given $\sigma_{B_l}$,  for any $\tau \in \{-1,1\}^t$ we have
\begin{align*}
\mu_n(\sigma_S = \tau |\mathcal{Y}_l = +) &= 
\sum_{\eta_l} \mu_n(\sigma_{B_l} = \eta_l|\mathcal{Y}_l = +) \mu_n(\sigma_S = \tau | \sigma_{B_l} = \eta_l)\\
&= O(\eps) + \sum_{\eta_l \in (\cup_{u \in S}U_\eps^l)^c} \mu_n(\sigma_{B_l} =\eta_l|\mathcal{Y}_l = +) \mu_n(\sigma_S = \tau | \sigma_{B_l} = \eta_l)\\
&= O(\eps) + \sum_{\eta_l \in (\cup_{u \in S}U_\eps^l)^c} \mu_n(\sigma_{B_l} =\eta_l|\mathcal{Y}_l = +)  Q_S^+(\tau) + O(\eps)\\
&= Q_S^+(\tau) \pm 3\eps.
\end{align*}

Since this holds for all $\eps > 0$, taking $\eps' = \eps/\max_\tau Q_S^+(\tau)$, we get that for any $\eps,t$, we can find large enough $n$, such that with probability $1-o(1)$ 

\[
\Pr_{\sigma\sim \mu_J}\big[\sigma_S=\tau \mid Y_\sigma = \pm\big]=(1\pm \epsilon)Q_S^{\pm}(\tau) = (1\pm \eps)Q_S^+(\tau)
\]
for all $\tau$. 

Finally, again using \eqref{eq:phase_lim} we get that the above holds if we replace $\mu_n(\sigma_{S} = \cdot|\mathcal{Y}_l = +)$ by $\mu_n(\sigma_{S} = \cdot|\mathcal{Y} = +)$.
From this, it follows that we can find a large enough $n$ and a large enough $l$ so that the required approximation of the third item holds.

It remains to establish \eqref{eq:limit}. Keeping $l$ fixed, since $\eta_l$ belongs in the $l$-hop neighborhood of $u$, we have by the local convergence of $\mu_n(\cdot|\mathcal{Y}_l = +)$ to $\nu^+$ that
\[
\lim_{n \to \infty} \mu_n(U_\eps^l(u)|\mathcal{Y}_l=+) = \Pr_{\eta \sim \nu^+}(U_\eps^l(u))
\]
where $\eta$ is sampled with law $\nu^+$ and $\eta_l$ is the set of leaves at distance $l$ from the root. 
Now, we also have that
\[
\limsup_{l \to \infty} \Pr_{\eta \sim \nu^+}(U_\eps^l(u)) \leq \Pr_{\eta \sim \nu^+}(\limsup_{l \to \infty} U_\eps^l(u)).
\]
Now notice that the event $\limsup_{l \to \infty} U_\eps^l(u)$ belongs to the tail $\sigma$-algebra of $\nu^+$ (where the tail is defined as $\cap_l \mathcal{F}_l$ and $\mathcal{F}_l$ is the $\sigma$-algebra generated by the spins at distance at least $l$ from the root).
Since $\nu^+$ is extremal, by applying \cite[Theorem 7.7(a)]{georgii2011gibbs} it follows that
\[
\Pr_{\eta \sim \nu^+}(\limsup_{l \to \infty} U_\eps^l(u)) \in \{0,1\}.
\]
Since $\eps > 0$, this probability cannot be $1$.
Hence, this tail event has zero probability 
and so we get \eqref{eq:limit}.
\end{proof}

\end{document}